\documentclass[11pt]{article}
\usepackage{newinutile-2.0}
\usepackage{multirow}
\usepackage{cite}

\usepackage{latexsym,amsmath,amsfonts,amsthm,amssymb,bbm,fdsymbol}
\usepackage{epsfig,graphics,color,calc,graphicx,pict2e}
\usepackage{multirow}
\usepackage[T1]{fontenc}
\usepackage[utf8]{inputenc}
\usepackage{multicol}
\usepackage{cite}
\usepackage{mathrsfs}
\usepackage{tikz}
\usepackage{tikz}
\usetikzlibrary{shapes,arrows}
\usetikzlibrary{intersections,matrix, positioning}

\usepackage{subfigure} 

\newlength{\pecettawidth}
\begin{document}
\title{\normalsize\Large\bfseries 
Statistical classification for Raman spectra of tumoral genomic DNA
}

\author[1]{Claudio Durastanti\thanks{claudio.durastanti@uniroma1.it}}
\author[1]{Emilio N.M. Cirillo\thanks{emilio.cirillo@uniroma1.it}}
\author[1]{Ilaria De Benedictis\thanks{ilaria.debenedictis@uniroma1.it}}
\author[3]{Mario Ledda\thanks{mario.ledda@cnr.it}}
\author[2]{Antonio Sciortino\thanks{antonio.sciortino@artov.immm.cnr.it}}
\author[3]{Antonella Lisi\thanks{antonella.lisi@cnr.it}}
\author[2]{Annalisa Convertino\thanks{annalisa.convertino@cnr.it}}
\author[2]{Valentina Mussi\thanks{valentina.mussi@cnr.it}}

\affil[1]{Dipartimento di Scienze di Base e Applicate per l'Ingegneria, 
             Sapienza Universit\`a di Roma, 
             via A.\ Scarpa 16, I--00161, Roma, Italy.}
\affil[2]{Institute for Microelectronics and Microsystems,
          CNR, via del Fosso del Cavaliere, 100, Roma, Italy.}
\affil[3]{Institute of Translational Pharmacology,
          CNR, via del Fosso del Cavaliere, 100, Roma, Italy.}

\date{\empty} 

\maketitle

\begin{abstract}
We exploit Surface--Enhanced Raman Scattering (SERS) 
to investigate aqueous droplets 
of genomic DNA deposited onto silver--coated silicon nanowires 
and
we show that it is possible to efficiently discriminate 
between spectra of tumoral and healthy cells.
To assess the robustness of the proposed technique, we develop 
two different statistical approaches, one based on the Principal 
Component Analysis of spectral data and one 
based on the computation of the $\ell^2$ distance between 
spectra. 
Both methods prove to be highly efficient and we test their 
accuracy via the so--called Cohen's $\kappa$ statistics. 
We show that the synergistic combination of the SERS spectroscopy and the  
statistical analysis methods leads to  
efficient 
and fast cancer diagnostic applications allowing 
a rapid and unexpansive discrimination between healthy and tumoral 
genomic DNA alternative to the more complex 
and expensive DNA sequencing. 
\end{abstract}







\vfill\eject
\section{Introduction}
\label{s:intro} 
\par\noindent
The relevance of Raman spectroscopy in medical diagnostics 
\cite{kksn2015}, and in particular in the study of cancer diseases
\cite{lpppsb2021}, has been widely pointed out in the recent 
pertinent literature. 
The potentiality of this technique lies in its label--free character, 
which allows to directly analyse biological samples and obtain unaltered 
information about their physico--chemical properties. 

Here, we exploited Raman spectroscopy to investigate aqueous droplets 
of genomic DNA deposited onto silver--coated silicon nanowires (Ag/SiNWs). 
By following the same experimental procedure proposed in \cite{mlpmpbllc2021}, 
we use Raman mapping to collect several spectra that are statistically 
analyzed here to discriminate between samples extracted from tumoral and 
healthy cells.

Raman spectroscopy is an inelastic optical scattering technique, 
which records the light scattered from vibrations  
in molecules or optical phonons in solids
\cite{mrr2007,tmrr2015}.
These inelastic scattering processes
have small cross section and, thus, 
the typical intensity of Raman signals is very low 
\cite{psp2003}.
On the other hand,
as it was firstly shown in \cite{fhm1974}, due to electromagnetic and 
chemical effects, the Raman signal coming from molecules adsorbed on a metal nanostructure
can be increased by several order of magnitude. 
This phenomenon, known as Surface--Enhanced Raman Scattering (SERS)
\cite{hmd2005,sdsd2008},
has been exploited in our experiment by dripping the DNA aqueous 
solution on a substrate made of a disordered array of
silver--coated silicon nanowires, whose 
effectiveness 
in enhancing the Raman signal has been recently 
demonstrated in a series of experimental studies, e.g.,
\cite{cmm2016,cmmllbfrl2018,zwlazc2008,gbcspb2009,zfzszwl2010,pcmmb2021,shb2012}.

In this way we are able to collect 
Raman maps of the deposited drops composed of several 
good quality Raman spectra 
which, although very similar among each others, 
allow us to distinguish between drops of healthy and tumoral DNA. 
Anyway the major challenge of this approach is that the Raman mapping  
generates large data sets that need an advanced data processing 
to extract meaningful information allowing us the discrimination between 
the healthy and tumor DNA sample. 

The aim of this work is thus to build a binary classification model 
aimed to discriminate, with 
high accuracy, spectra coming from different DNA molecules, corresponding to 
the outcomes of a target variable taking values 0 and 1 for healthy and 
tumoral samples respectively. 

Our proposal is based on two different methods.  
The first one reduces initially the amount of 
predictors by means of a Principal Components 
Analysis (PCA). After reducing dimensionality, 
the new variables are exploited to build a logistic 
regression model, whose outcome is the desired classifier. 
The second method involves the full data set to exploit the 
geometric features of 
the Raman spectra. Indeed the classifier adopted in this case 
is based on the computation of the $\ell^2$ distance 
between test samples and the spectral average of healthy and tumoral 
training spectra. We will show that both the strategies achieve a very 
high accuracy, close to 90\%. 

The paper is organized as follows: in Section~\ref{s:matmet}
we discuss experimental and statistical methods, mainly focusing on 
the latter. Our results are discussed in Section~\ref{s:risultati}, 
whereas in Section~\ref{s:conclusioni} we summarize our 
conclusions. 

\vfill\eject
\section{Methods}
\label{s:matmet} 
\par\noindent
In this section we describe the 
processes and approaches that we have 
developed in the different steps of our study.
We shall provide a very short account of the sample preparation 
procedure and the Raman measurements, referring to 
\cite{mlpmpbllc2021} for a thorough and detailed description. 
On the other hand, we shall discuss in detail the statistical 
approach developed to analyze the experimental data, 
which is the true novel contribution provided in this paper.

\subsection{Experimental procedures}
\label{s:sperimentale} 
\par\noindent
Plasma enhanced chemical vapor deposition (PECVD) was used to grow 
Au catalyzed SiNWs on Si wafers, kept at $350^\textup{o}$C,
using SiH$_4$ and H$_2$ as precursors. 
The coating was realized by evaporating an Ag film onto SiNWs 
arrays with nominal thickness of $100$~nm.

\begin{figure*}[t]
\centering
\includegraphics[width=.4\columnwidth]{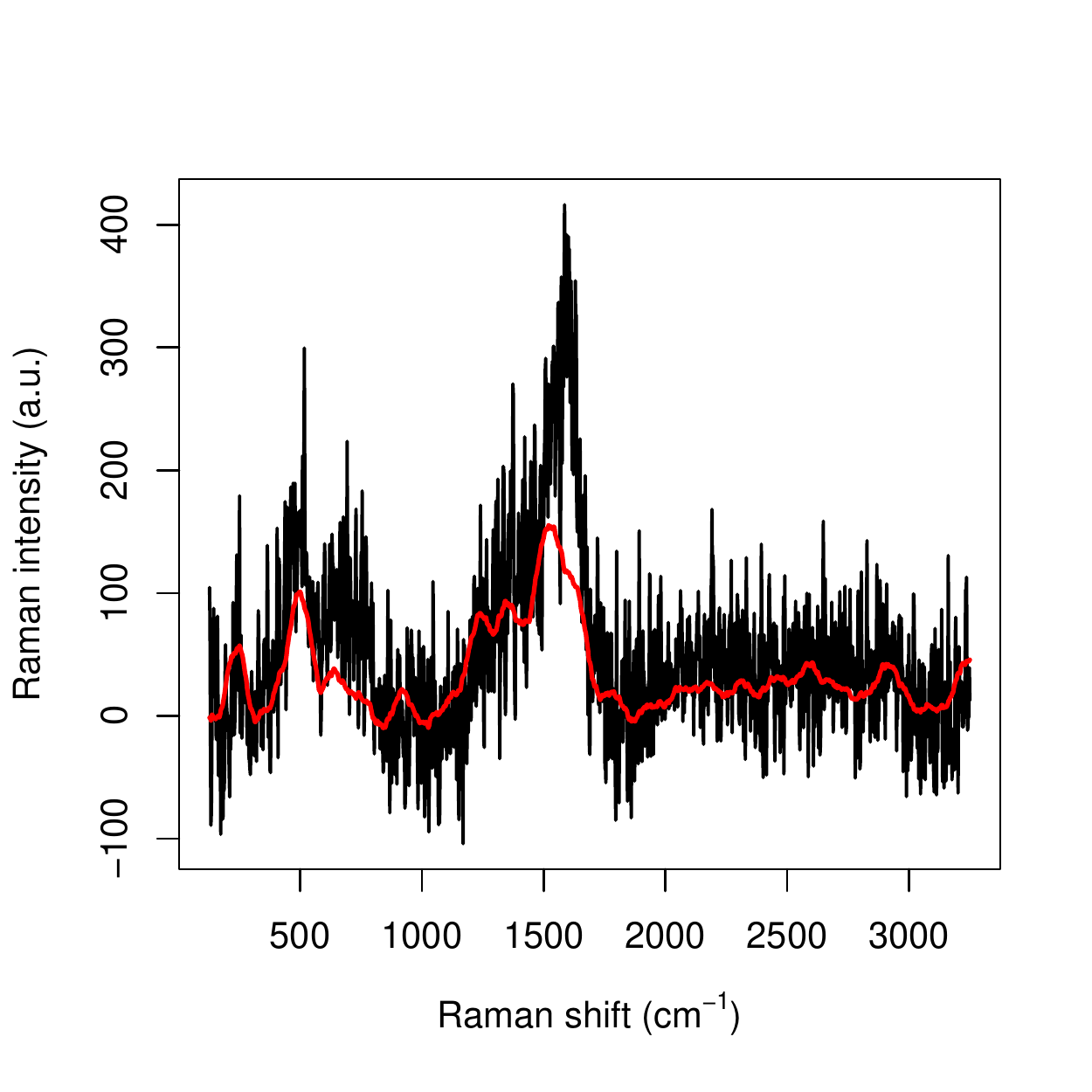}
\caption{Comparison between a raw Raman spectrum and the curve (in red) 
obtained by 
performing Savitzky--Golay filtering.}
\label{f:fig01}
\end{figure*}

As skin cancer model \cite{wlhwgkhoe2016} we used the human melanoma cell line 
SK--MEL--28, which was compared to the human immortalized 
keratinocyte HaCaT as control health skin model.
After standard culture, harvesting, and centrifugation, cell pellets 
were obtained and used to extract the genomic DNA, that, on turn, 
was re--suspended in DNase free water to obtain a $20$~ng$/\mu$L solution.

The final samples are prepared by depositing one drop of the DNA 
solution on Ag/SiNWs substrates coming from the very same batch.

Each droplet is
spectrally mapped after drying by means of a DXR2xi Thermo Fisher Scientific 
Raman Imaging Microscope equipped with a
$532$~nm excitation laser 
and a $50\times$ objective.
For each droplet, Raman spectra are collected at points on a square 
grid with spacing $4~\mu$m,
at $1$~mW laser power and 
performing four accumulations lasting
$5$~ms each.

\vfill\eject
\subsection{Data set and conveyed information}
\label{s:dataset} 
\par\noindent
As data set we have considered $N=3980$ spectra, 1990 from the 
tumoral and 1990 from the healthy DNA droplet. 
The spectra have 
been randomly chosen in the central part of the 
droplets, 
since in this region the biological layer formed after dehydration 
happens to be thinner, so that the interaction between the single 
molecules and the nanostructured substrate is direct and tighter, 
and the SERS effect more effective, ensuring a good quality of 
the Raman spectra. Due to the fact that the spectra are collected 
at points of the droplet at distance larger than, or equal to, 
$4~\mu$m while the size of a DNA molecule is of nanometer order, 
we can reasonably assume that our data are independent. 
A typical Raman spectrum we will deal with is reported in 
figure~\ref{f:fig01}. In order to explain its 
main features we have to review some basic facts about Raman 
spectroscopy. 
In Raman spectroscopy incident photons are scattered by molecules
in such a way that
the emitted photon has energy different from that of the incident one
and molecules, after the interaction, jump to a different 
vibrational energetic level. This phenomenon is often explained
saying that the molecule absorbs the incident photon and jumps to a 
``virtual" energy level with an incredibly short life--time
(say less than $10^{-15}$~sec); then it relaxes to a vibrational energy 
state different from the initial one emitting a photon which, 
consequently,
has energy different from that of the incident one.
The total emitted energy, suitably normalized and expressed in arbitrary 
units, is called 
\emph{Raman intensity}, whereas the energy difference between the 
incident (laser) light and the scattered (detected) light,
expressed in cm$^{-1}$, is called \emph{Raman shift}.

Raman spectroscopy is a powerful method in investigating 
biological systems 
because it provides a
molecular fingerprint of the samples 
in a completely label--free way and with high specificity 
\cite{kksn2015}. 
Indeed, the peaks appearing in the spectra are 
uniquely associated with particular chemical structure 
present in the molecule.
For instance, 
in the raw spectrum reported in figure~\ref{f:fig01}, where no 
pre--processing has been performed on the data obtained from 
the spectrometer, some peaks are perfectly visible and can 
be roughly located at 
$230$~cm$^{-1}$, 
$540$~cm$^{-1}$, 
$1320$~cm$^{-1}$, 
$1570$~cm$^{-1}$, 
and
$2930$~cm$^{-1}$.
Very precise measures of the peak positions 
can be performed by suitably averaging several spectra, see, e.g.,   
\cite[Supplementary material, Fig.~S5]{mlpmpbllc2021},
but we note that even in the data 
reported in figure~\ref{f:fig01} the CH--group vibration peak 
at $2930$~cm$^{-1}$ 
and the Ag--N stretching vibration mode 
at $230$~cm$^{-1}$ 
are perfectly visible.

As regards the diagnostic potential of Raman spectrosocopy, 
cancer is nowadays associated with changes in the 
nucleotide sequence of DNA molecules \cite{scf2009} which also induces
variations in DNA physical properties, such as stiffness, length, and shape. 
It has been demonstrated in \cite{mlpmpbllc2021}, 
in our experimental set--up, that
the variation of these physical properties causes modifications in the 
interaction between DNA molecules and the nanostructured substrate 
which reflect in the observed Raman spectra. 

\subsection{Data pre--processing}
\label{s:pre} 
\par\noindent
Each Raman spectrum here considered consists in $p=1680$ Raman intensity 
values corresponding to $p$ values of the Raman shift, lying in the 
interval between $50.6$ and $3288.5$~cm$^{-1}$. 
Moreover, we obtained smoothed 
data by filtering 
the original raw spectra with 
the Savitzky--Golay algorithm \cite{SG64} (see also \cite{ZK13})
over a window of $90$ data points treated as convolution coefficients.
In figure~\ref{f:fig01} we plotted a raw spectrum and 
its smoothed version.

Although, the large part of the collected spectra share a similar behavior, 
there are some that are suspiciously different from the others.
These spectra have been considered as 
outliers associated to local experimental fluctuations and thus eliminated 
from the analysis.

We build a decision surface to identify and remove outliers from both 
the healthy and tumoral
data sets by adding and subtracting three times the (point--wise) empirical 
standard deviations to the average spectra. All the spectral patterns 
featuring at least a point out of the decision surface are then discarded.

\begin{figure*}[t]
	\centering
	\includegraphics[width=.45\columnwidth]{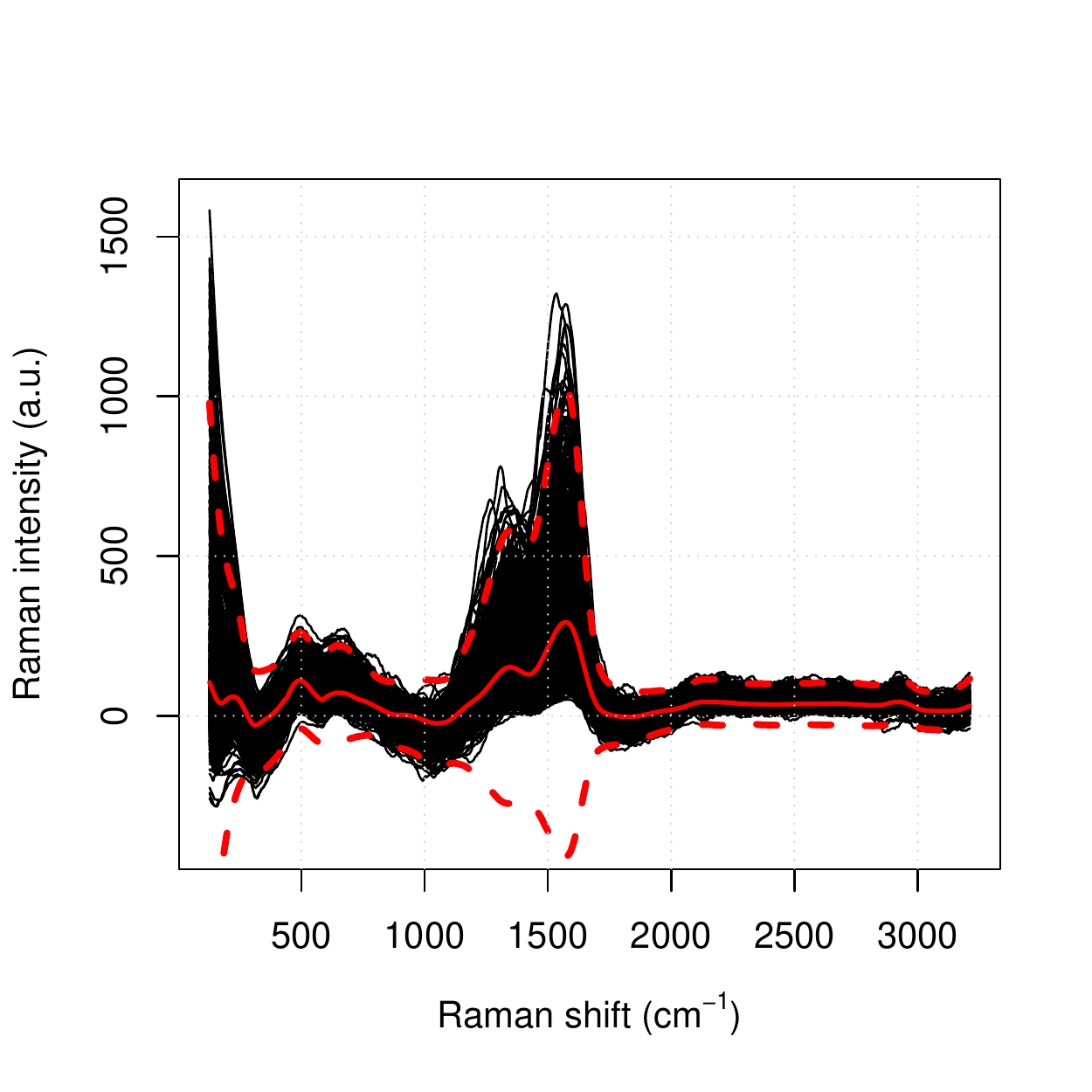}
	\hskip 1. cm
	\includegraphics[width=.45\columnwidth]{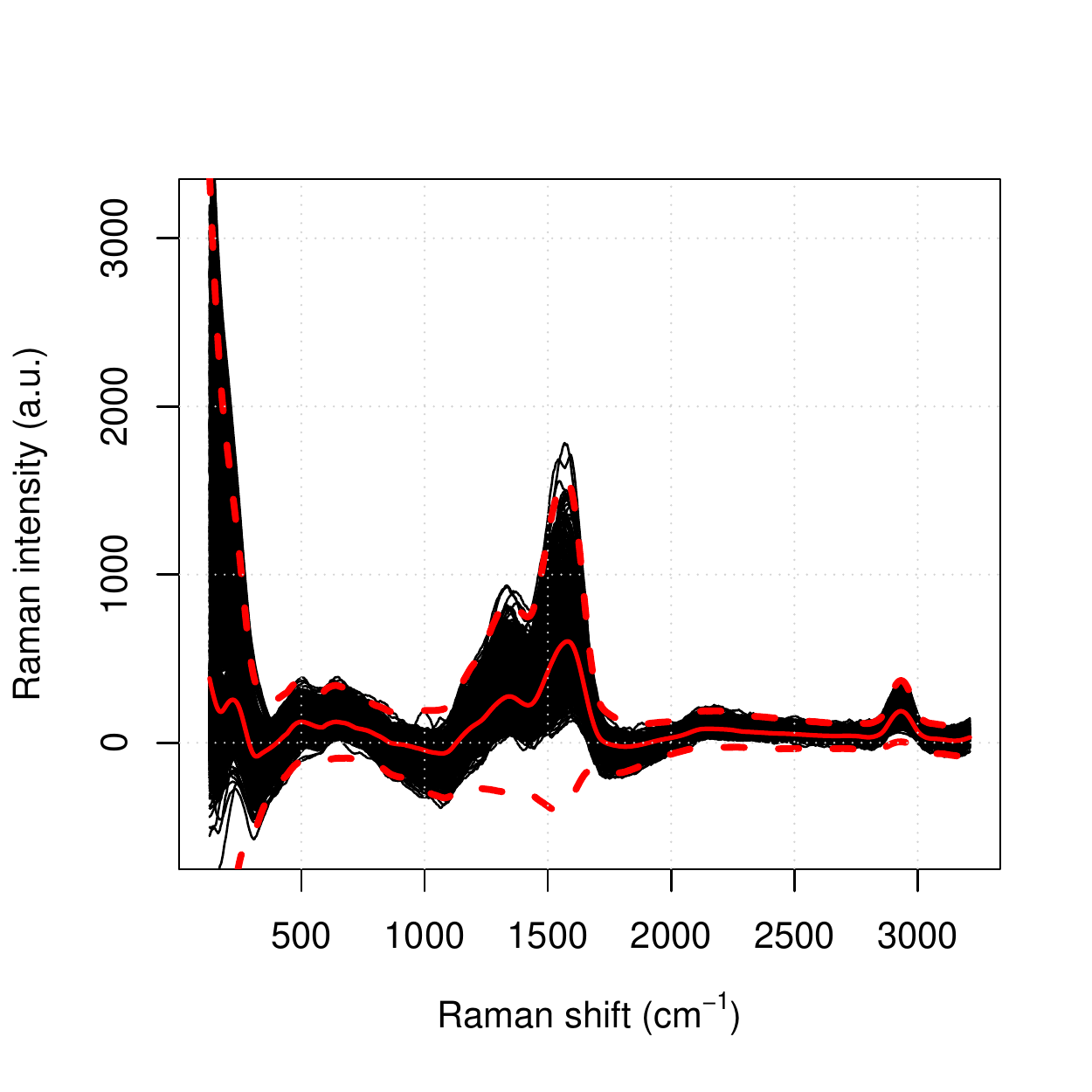}
	\caption{For the 
		healthy (left) and tumoral (right) data set we 
		report the smoothed spectra (black solid lines), 
		the average spectra (red solid lines), and 
		the extreme curves computed by 
		adding to and subtracting from the average Raman intensity three times 
		the standard deviation (red dashed lines) labeling the decision surfaces.}
	\label{f:fig02}
\end{figure*}

Figure~\ref{f:fig02} shows the whole set of spectra for the healthy and tumoral cell, in the left and in the right panel  respectively. 
Both the panels show the corresponding decision surfaces, labeled by average spectra (solid lines) and the extreme curves (dashed lines). 
In both the cases, the selection procedure allows us to discard 
approximately 15\% of the available spectra. 

\subsection{Two statistical approaches}
\noindent 
We will propose two different models, one based on the 
PCA analysis and one based on the computation of $\ell^2$ distance. 
The spectra are split into test and training sets, 
the first used to tune simultaneously the parameters of both the models here proposed, the second to validate them.   
As aforementioned, we deal with a binary target  variable $W$, whose outcomes 1 and 0 correspond to tumoral and healthy DNA molecules respectively.

\subsubsection{The local method: PCA analysis and logistic regression}
\label{s:pca} 
\par\noindent
Borrowing the notation from \cite[Chapter~1 and Paragraph~8.2.1]{htw2015}, 
any single spectrum is represented as a column vector $x\in\mathbb{R}^p$.
By collecting the $N$ row vectors $x^\dag$, where $\dag$ denotes transpositions, we construct the 
$N\times p$ matrix $\mathbf{X}$ which represents the entire data set. 
The $j$--th column of $\mathbf{X}$ is 
the collection of the $N$ observations of the $j$--th variable, 
namely, the intensity corresponding to the $j$--th value of the Raman shift. 

Thus, we produce the $N\times p$ matrix $\mathbf{Y}$ by centering $\mathbf{X}$ with respect to the columns (i.e., the Raman shift). 
More in details, each entry $y_{ij}$ of $\mathbf{Y}$ is computed by subtracting 
to each entry $x_{ij}$ of $\mathbf{X}$ 
the sample mean computed along the elements corresponding to the same Raman shift, 
that is to say by setting 
\begin{equation}
\label{pca000}
y_{ij}
=
x_{ij}-\frac{1}{N}\sum_{s=1}^N x_{sj}
\end{equation}
for any $i=1,\dots,N$ and $j=1,\dots,p$. 
A \emph{principal components analysis} is then obtained by the  eigendecomposition of the  empirical covariance matrix $\textbf{Y}^\dag\textbf{Y}$  (see, for example, \cite{Jackson91}). The so--called  \emph{principal components directions}
$v_1,\dots,v_p\in\mathbb{R}^p$ are computed and, as usual, we call $i$--th \emph{principal component loadings}
the $p$ elements 
of the column vector $v_i$.
The projection $\mathbf{z}_i=\mathbf{Y}v_i\in\mathbb{R}^N$ is called $i$--th 
\emph{principal component} of the data $\mathbf{Y}$. 
The variance of each principal component (PC) is given by the corresponding 
eigenvalue and  it concentrates on the 
first $m$ principal components, allowing us to neglect in the next step all the other $p-m$ components.
 
The selected first $m$ principal components $z_1,\dots,z_m$ are 
thus interpolated to build a logistic regression model to estimate 
the  probability mass function of the binary 
target variable $W$ by 
\begin{equation}
	\label{pca020}
\textup{Pr}(W=1|z_1,\dots,z_m)=\frac{e^{\beta_0+\sum_{i=1}^m\beta_iz_i}}
                                    {1+e^{\beta_0+\sum_{i=1}^m\beta_iz_i}}
\;\;\textup{  and }\;\;
\textup{Pr}(W=0|z_1,\dots,z_m)=\frac{1}{1+e^{\beta_0+\sum_{i=1}^m\beta_iz_i}},
\end{equation}
where $\beta_i\in\mathbb{R}$ for $i=0,\dots,m$.
In this case, we consider an optimization parameter $\lambda\in[0,1]$ such that we associate the outcome for the binary variable $W=1$ to each set of components $z_1,\dots,z_m$ if  $\textup{Pr}(W=1|z_1,\dots,z_m)\ge\lambda$ and $W=0$ otherwise. 
This method is referred as ``local'' since the PCA analysis selects 
a subset of spectral indices to be the most relevant within the subsequent 
regression.

\begin{figure*}
	\centering
	\includegraphics[width=.45\columnwidth]{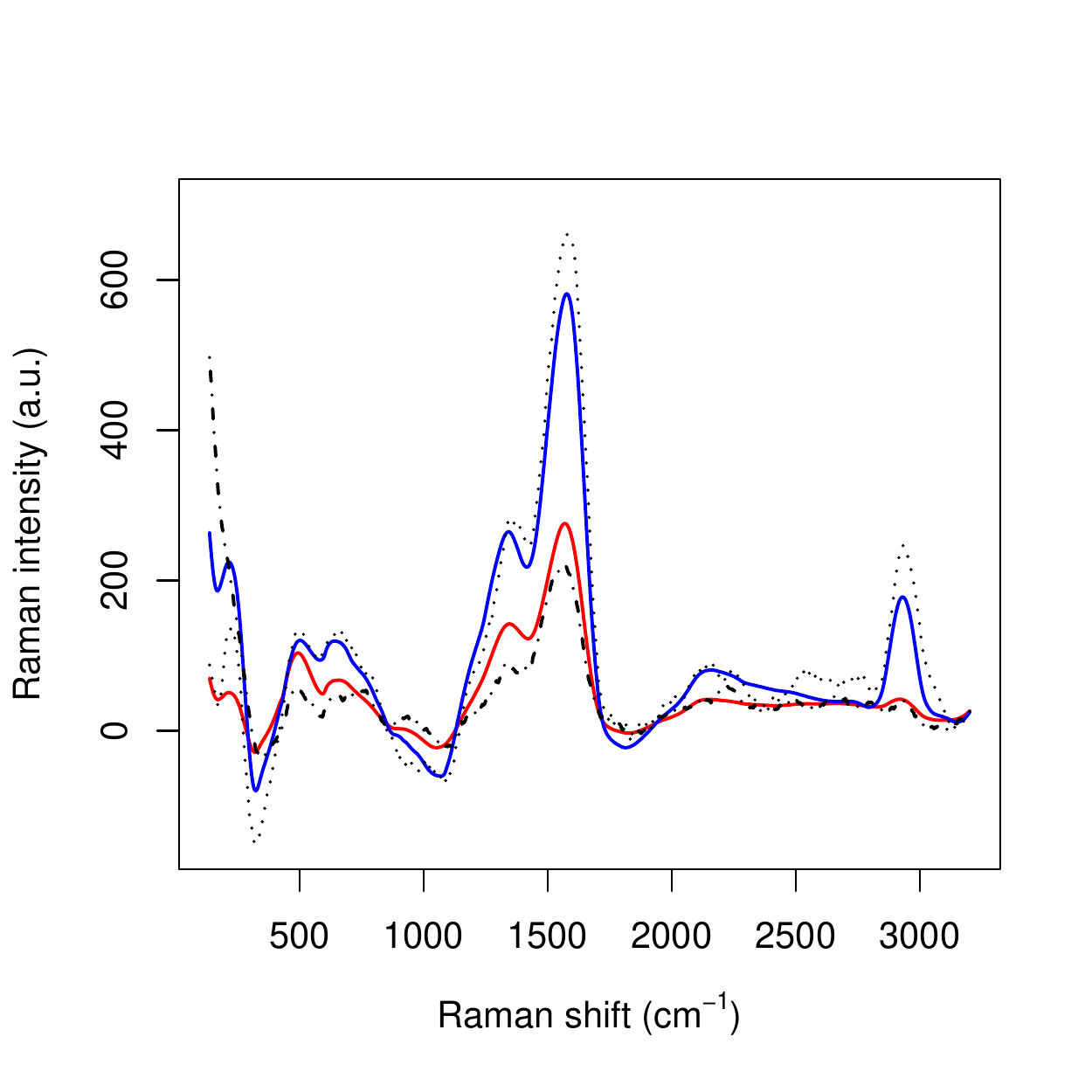}
	\caption{
                The red and the blue solid lines are, respectively, 
		the average spectrum of the healthy and the tumoral data set. 
                The point--dashed and the dashed lines report, respectively, 
                one healthy and tumoral pre--processed spectrum. 
                For the healthy spectrum 
		$\textup{d}_t=14.9\times 10^6$ 
		and 
		$\textup{d}_h=4.7\times 10^6$.
                For the tumoral spectrum 
		$\textup{d}_t=2.8\times 10^6$ 
		and 
		$\textup{d}_h=16.6\times 10^6$.
	}
	\label{f:fig03}
\end{figure*}

\subsection{The global method: geometric analysis}
\label{s:geo} 
\par\noindent
The second proposed strategy aims to distinguish spectra coming from the healthy and tumoral data set following their geometric features.

After splitting the dataset into training and test sets, the training set is used to produce 
the healthy and the tumoral average spectra, represented by the 
column vectors $h$ and $t$ of $\mathbb{R}^p$.
Thus, for each spectrum belonging to the test set represented by the 
$i$--th row of the data matrix $\mathbf{X}$, we 
compute the $\ell^2$ distances 
\begin{equation}
	\label{geo000}
	\textup{d}_h(i)
	=
	\sum_{s=1}^p|x_{is}-h_s|^2
	\;\;
	\textup{ and }
	\;\;
	\textup{d}_t(i)
	=
	\sum_{s=1}^p|x_{is}-t_s|^2
	.
\end{equation}
so that each spectrum can be classified by setting the following outcome binary function
\begin{equation}
	\label{geo020}
	g_\textup{out}(i)
	=
	\mathbbm{1}\{\tau \textup{d}_t(i)\le(1-\tau)\textup{d}_h(i)\}
	,
\end{equation}
where $\tau\in[0,1]$ is an optimization parameter. 
As above, the outcome 
is equal to $0$ if the spectrum is identified as coming from healthy DNA molecules and 
equal to $1$ otherwise. 
This method is referred as ``global'' since the $\ell^2$ distance is 
computed over the complete set of spectral data.

\begin{figure*}[t]
	\centering
	\includegraphics[width=.45\columnwidth]{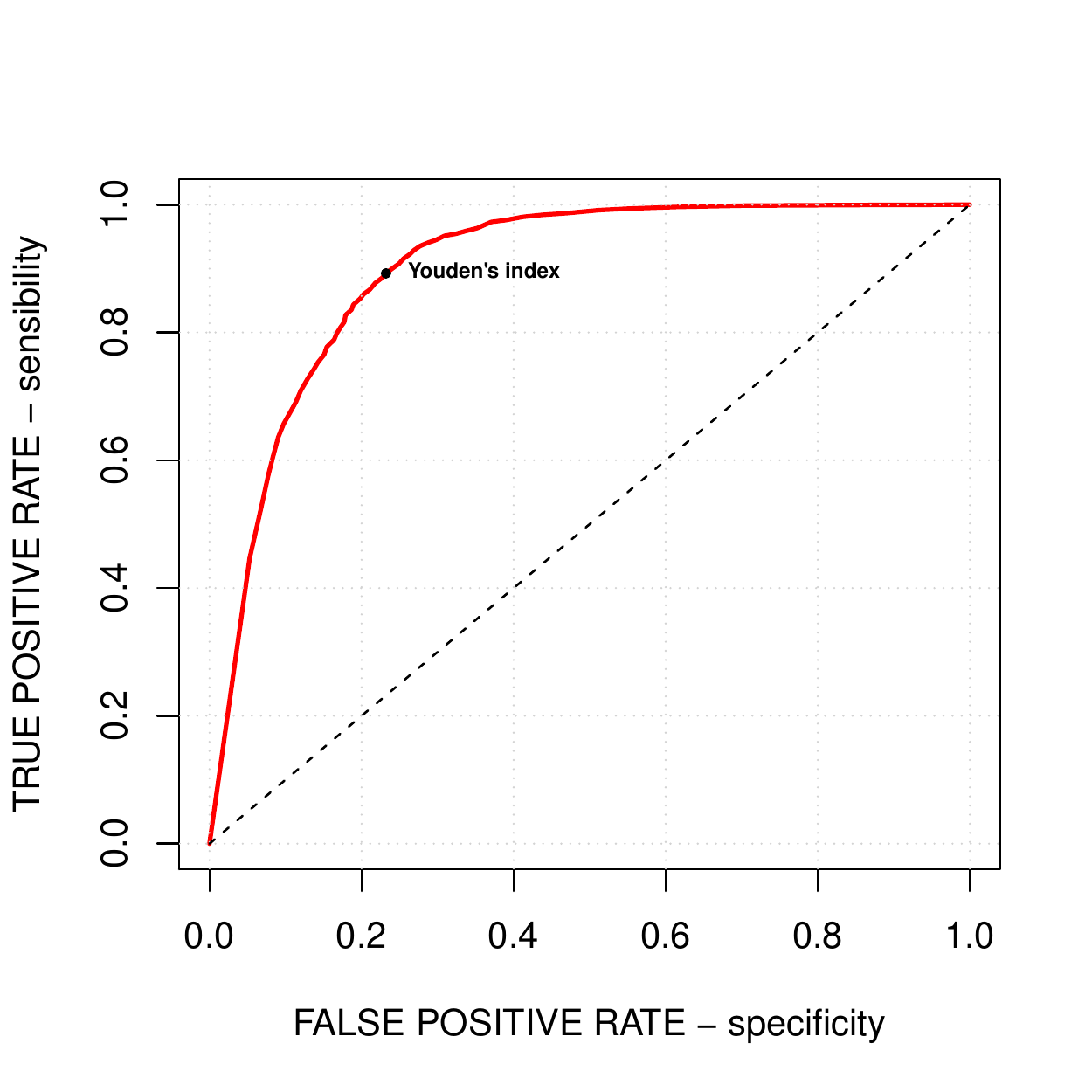}
	\hskip 1. cm
	\includegraphics[width=.45\columnwidth]{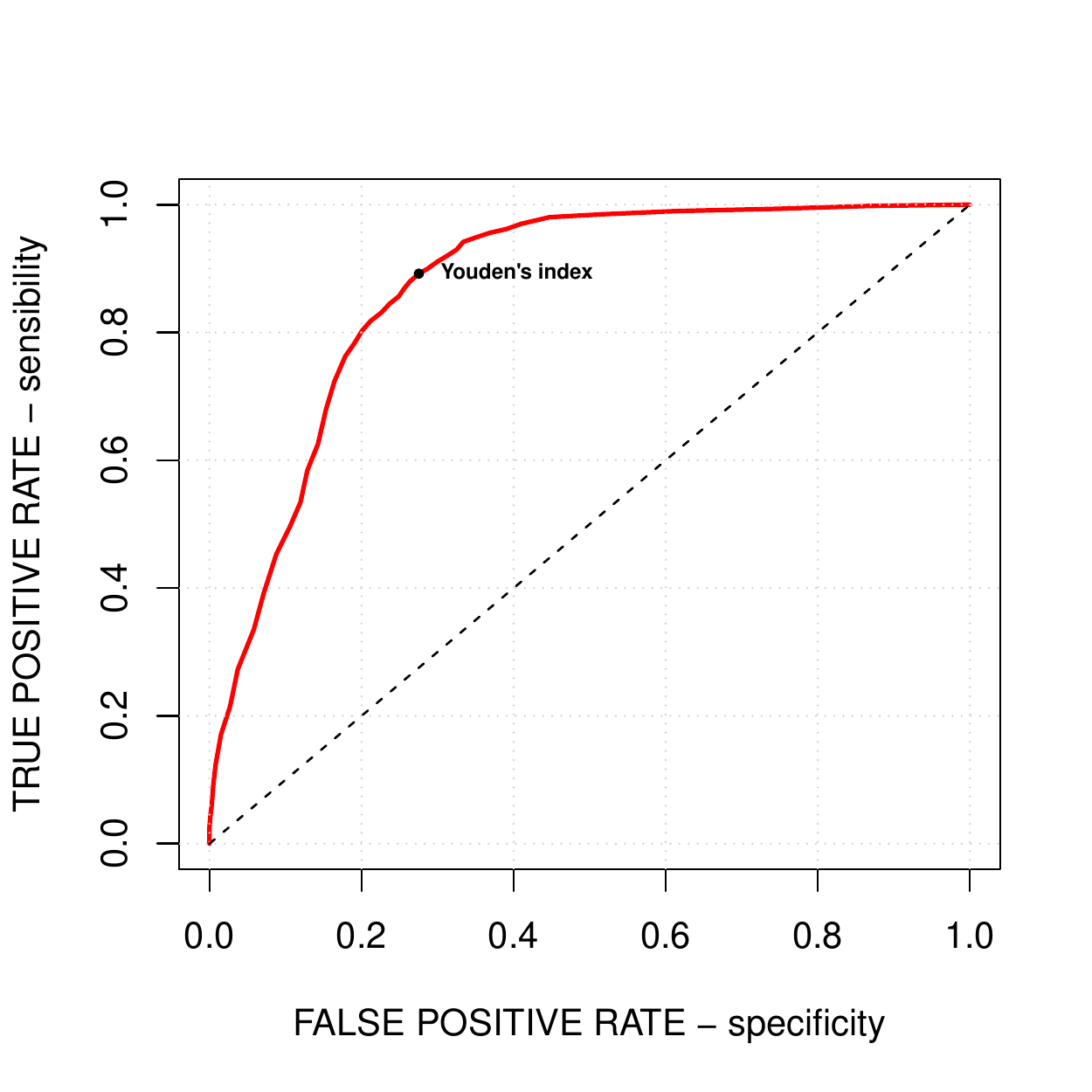}
	\caption{ROC curves for the local (left panel) and 
the global (right panel) methods.
The markers put in evidence the corresponding Youden's indices.}
	\label{f:fig04}
\end{figure*}

\begin{figure*}[t]
	\begin{picture}(80,180)(-10,0)
        \put(0,0){
	\includegraphics[width=.45\columnwidth]{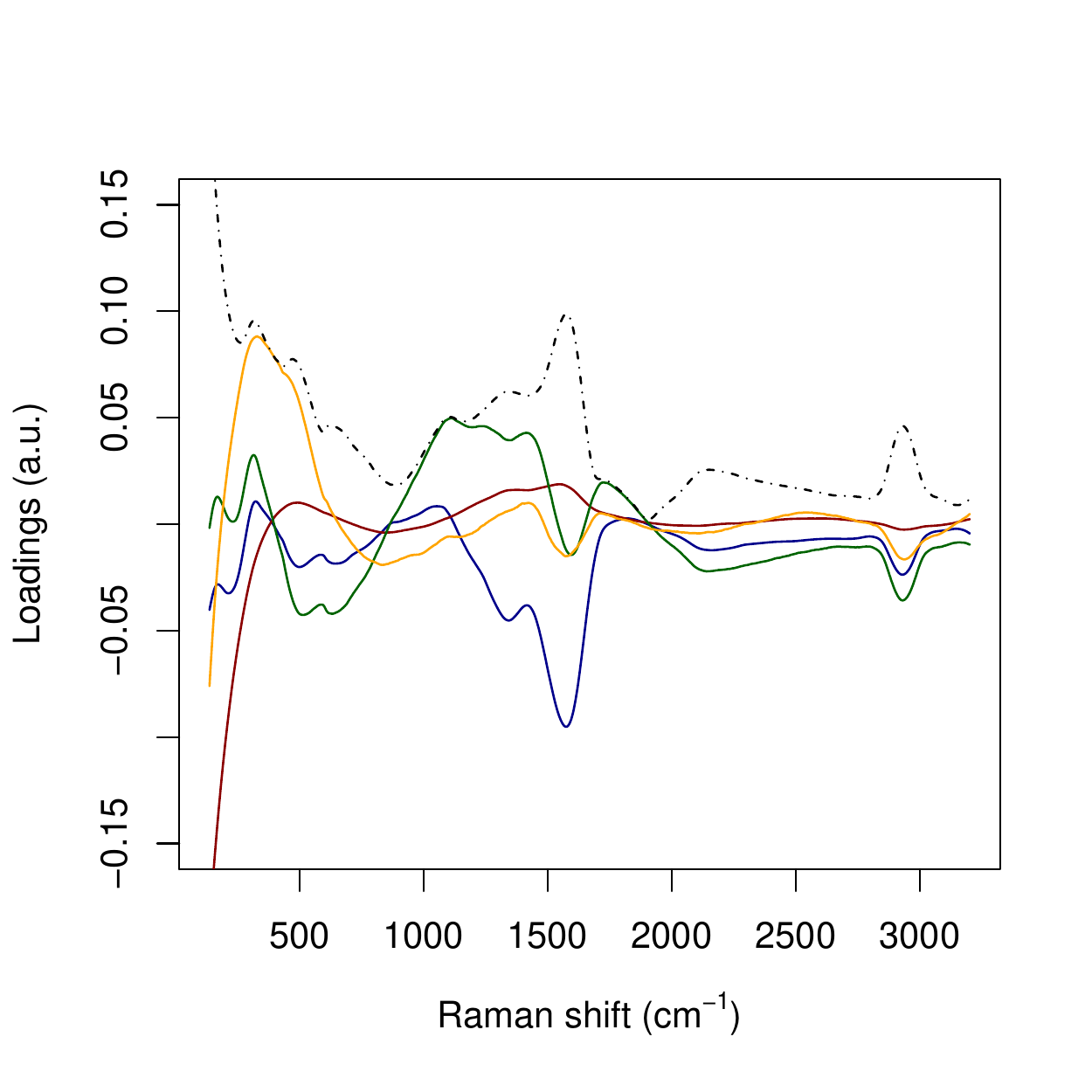}
        }
        \put(240,15){
	\includegraphics[width=.39\columnwidth]{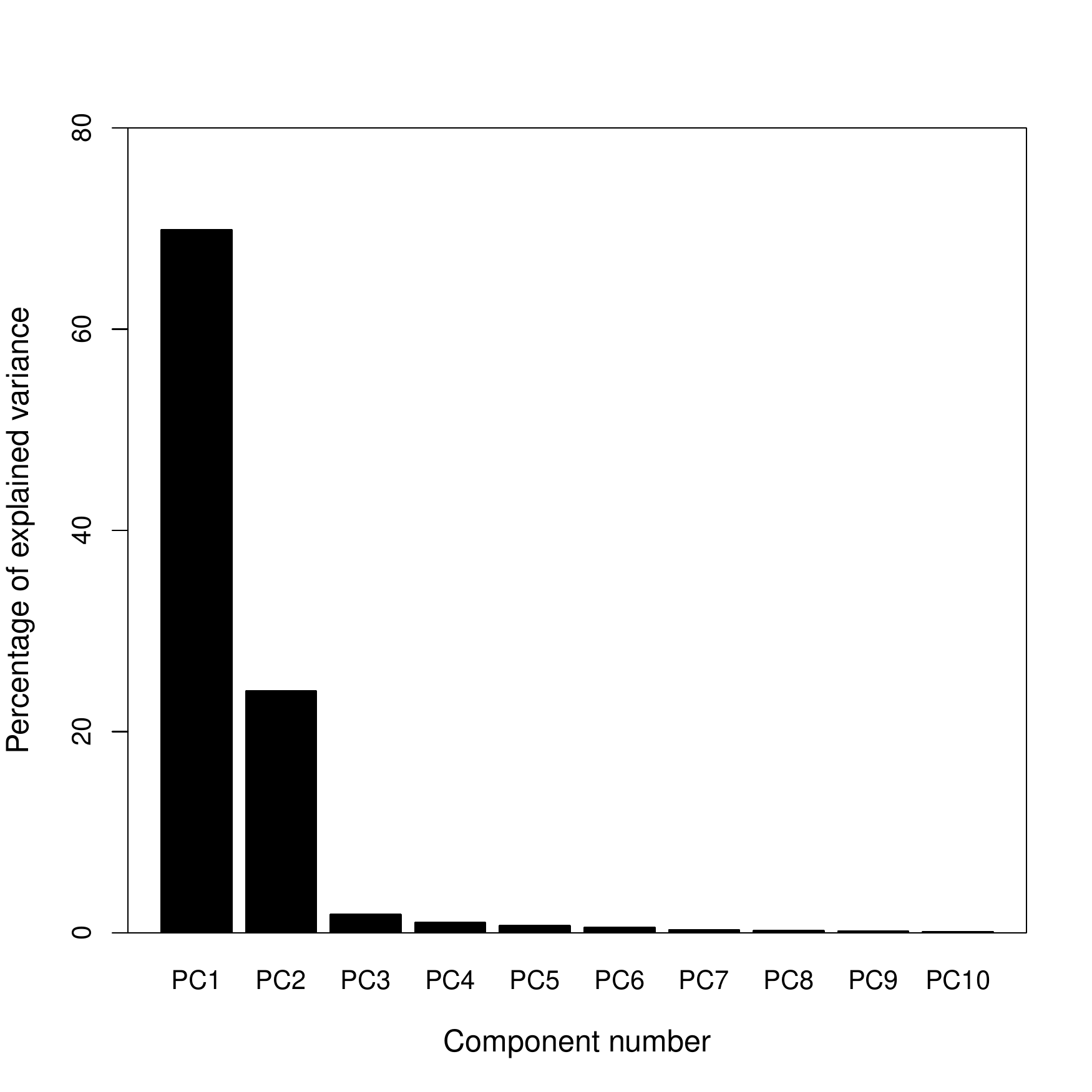}
        }
	\end{picture}
	\caption{Left: the four solid lines reports the
		loadings associated to the first four principal components 
                (blue first, brown second, green third, and yellow fourth); 
                the dashed black lines is 
		the intensity, i.e., the square root of the sum 
                of the squares, of the loadings of the first four components. 
                Right:
		percentage of variance as function of the principal component index.
		}
	\label{f:fig05}
\end{figure*}

\begin{figure*}[t]
	\centering
	\includegraphics[width=.6\columnwidth]{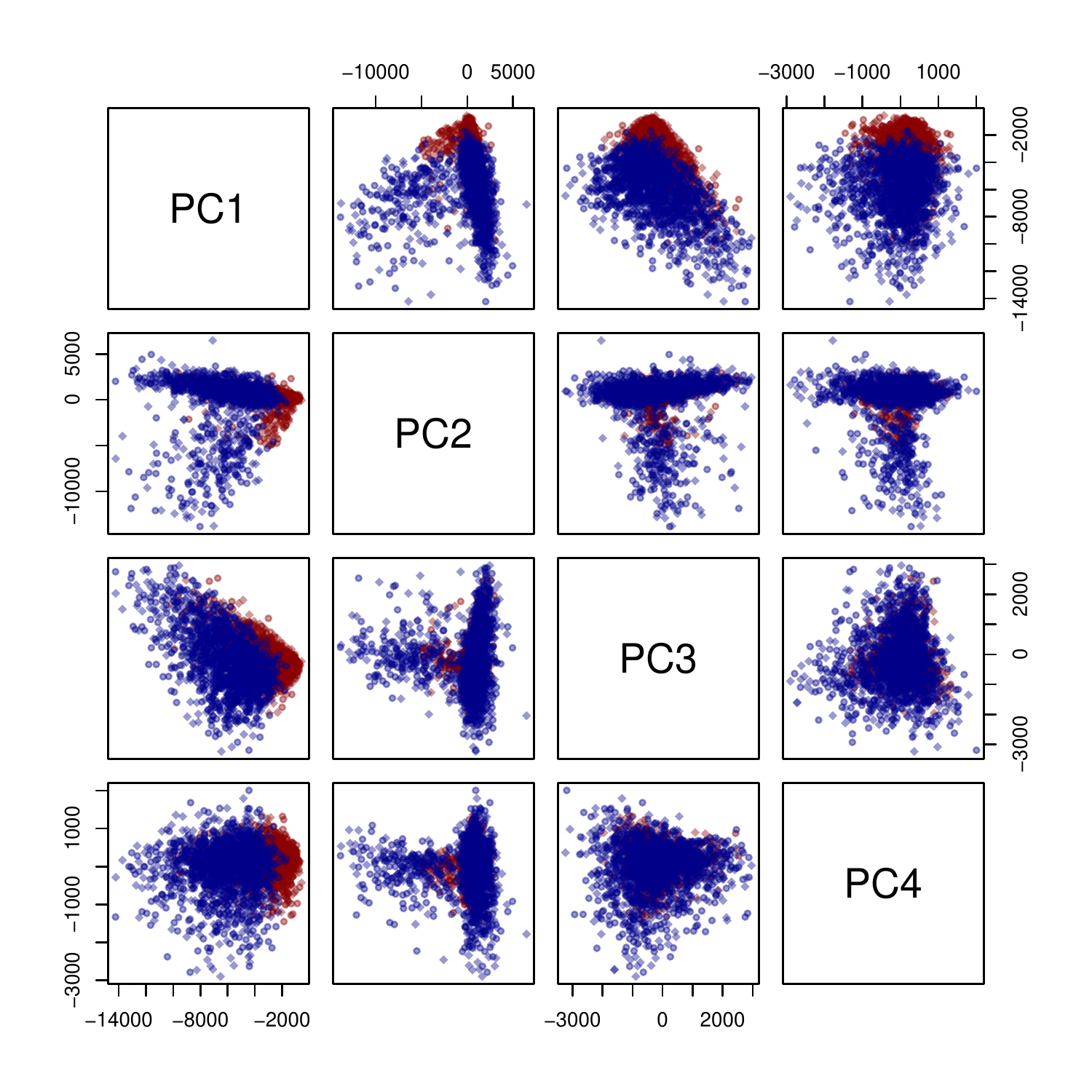}
	\caption{
		Projections on the coordinate planes of the distribution of the first four principal components on the coordinate planes. 
		}
	\label{f:fig06}
\end{figure*}

\begin{table}
\centering
	\begin{tabular}{ |c||c|c| }   	\hline
		Method & optimal tuning parameter & Area Under the Curve \\
		\hline
	local   & $\lambda=0.46$ &0.899 \\
		global   & $\tau=0.34$& 0.871\\
		\hline	
	\end{tabular}
	\caption{Optimal tuning parameters and AUC 
                 for both the methods.
                 }
        \label{t:ROC}
\end{table}

\begin{table}
\begin{tabular}{ |c||c|c|c| }   	\hline
		Principal component & Standard deviation & Proportion of Variance & Cumulative Proportion \\
	\hline
	PC1   &  5157.11860 & 0.78068 & 0.78068 \\
	PC2   &  2311.49810 & 0.15684 & 0.93752 \\
	PC3   &   866.34106 & 0.02203 & 0.95955 \\
	PC4   &   581.91444 & 0.00994 & 0.96949 \\
	PC5   &   496.66847 & 0.00724 & 0.97673 \\
	PC6   &   405.79667 & 0.00483 & 0.98156 \\
	PC7   &   327.32049 & 0.00314 & 0.98471 \\
	PC8   &   284.63234 & 0.00238 & 0.98708 \\	
	\hline	
\end{tabular}
	 \caption{Variance analysis (first 8 principal components).
          	\label{t:tab01}}
\end{table}

\section{Results and discussion}
\label{s:risultati} 
\par\noindent 
We estimate first the tuning parameters $\lambda$ and $\tau$ by means of a 
$10$--fold cross--validation procedure, a popular method used to
evaluate predictive models over a limited amount of sampled 
data (see \cite[Ch. 7]{HTF09}). 

The original sample is randomly partitioned into 10 equally sized subsamples. A subsample is kept as test set, while the other 
nine ones are used  as training data. Then the accuracy of both the methods are evaluated,  while the cross--validation process is repeated 10 times, paying attention to use each round a different subsample as test group. The 10 results are then averaged to compute a single estimation. 
The advantage of this validation strategy is that all observations are
used at the same time for both training
and testing and each observation is used for testing exactly once. 

The accuracy of the choice of the tuning parameters is evaluated by means of the so--called Youden's $J$ statistic, or -- simpler, Youden's index, given by the formula 
\begin{equation}
\label{res000}
	\text{Youden's index} = \text{sensitivity}+\text{specificity}-1, 
\end{equation}
where sensitivity and specificity are the true and false positive rates respectively. 
Figure~\ref{f:fig04} explores the trade--off between specificity and sensitivity by means of the corresponding ROC curve, while Table~\ref{t:ROC} presents for both the models the choice of the optimal tuning parameters, corresponding to the maximal Youden's indices, and the Area Under the Curve (AUC) value, measuring the two--dimensional area under the ROC curve and providing thus an aggregate measure of the performance across all possible choices of the tuning parameters (see, for example, \cite{Fawcett2006}).

As shown in Table~\ref{t:ROC},
both $\lambda$ and $\tau$ are smaller than $1/2$, which is the exact balance 
of the healthy and tumoral subsets. 
However, it matches our goals 
since these values of the tuning parameters
reduce the amount of false negatives (i.e., not detecting tumoral 
samples) paying the price of 
a larger amount of false positives (i.e., misclassifying healthy samples), 
which seems to be the fairest choice.  

Concerning the PCA analysis, as shown in Table~\ref{t:tab01} 
(see the right panel in figure~\ref{f:fig05}), 
$97.0\%$ circa of the variance concentrates on the 
first $m=4$ principal components. 
Furthermore, figure~\ref{f:fig06} 
shows a strong separation between the third and  fourth principal component. 
Thus, from now on, only the first 4 principal components are selected. 
Also, each discarded component is characterized by a proportion of 
variance sensibly smaller than $1\%$ (see again Table~\ref{t:tab01}).

\begin{table}
	\begin{center}
	\begin{tabular}{ |c||c|c| }   	\hline
		$i$ & Estimate & Standard Deviation\\
		\hline
		0   &  -6.6094526 & 0.2424116 \\
		1   &  -0.0014903 & 0.0000540 \\
		2   &  -0.0002585 & 0.0000440 \\
		3   &  -0.0020081 & 0.0001042 \\
		4   &  -0.0007470 & 0.0001154 \\
		\hline	
	\end{tabular}
	\caption{Estimates of $\beta_i$ and corresponding errors.
	\label{t:tab02}}
	\end{center}
\end{table}

The left panel in figure~\ref{f:fig05} collects the loadings, that is, the coefficients of the linear combinations of the original variables defining the principal components, related to the first four PC's, and, thus, provide a measure for the contribution of each observable to the main components. 
The solid lines are concerned with the first four PC's separately, 
while the dashed black line represents the associated intensity, that is, 
the $\ell^2$--norm associated to the loadings of the considered PC's. 
The three main peaks appearing in the panel correspond to 
wavenumber 230 cm$^{-1}$, 
$1550$ cm$^{-1}$, 
and 2930 cm$^{-1}$.
The first one is related to the Ag--N stretching vibration mode.
The second one is contributed both by the overlapping of C--C and C--N 
stretching vibrations involving the aromatic rings of the DNA bases, 
and the so--called ``cathedral peaks", deriving from the formation of a 
carboneous layer due to the photo--decomposition of the DNA at the silver 
surface upon laser irradiation.
The last one is
linked to the vibrations of the CH--group. 

The estimates for the coefficients $\beta_i$, $i=0,\ldots,4$ 
obtained by the logistic regression are collected in Table~\ref{t:tab02}.
Recall that, in this case, the optimization parameter $\lambda\in[0,1]$ is defined such that $W=1$ for each $z_1,\dots,z_4$ 
if $\textup{Pr}(W=1|z_1,\dots,z_4)\ge\lambda$, otherwise $W=0$. 

\begin{table}
	\begin{center}
		\begin{tabular}{ |c||c|c| }   	\hline
			Population (col.)  vs. Prediction (row)& 
                                      Positive (\%)& Negative (\%)\\
			\hline
			Positive (\%)  & 44.6/44.6 & 11.6/13.8 \\
			\hline
			Negative (\%)  & 5.3/5.4 & 38.4/36.2 \\
			\hline	
		\end{tabular}
\caption{Confusion matrix (local method / global method).
         \label{t:CMPCA}
}
	\end{center}
\end{table}

We evaluate, now, 
the accuracy of the proposed methods by means of the 
two confusion matrices in 
Table~\ref{t:CMPCA}
(see, again, \cite{Fawcett2006})
computed by averaging 
true positives (TP), true negatives (TN), false positives (FP), 
and false negatives (FN)
over all the possible choices of the training and the test set of data.
We use the values collected 
in 
Tables~\ref{t:CMPCA} 
to calculate, for both 
the methods, the 
so--called Cohen's $\kappa$ statistic, which takes values 
in $[-1,+1]$ and measures the 
agreement between two classifiers and can 
also be used to assess the performance of a classification model, 
given by the formula (in the binary case) 
\begin{equation}
\label{res010}
	\kappa=\frac{2\left( \text{TP}\times\text{TN}-\text{FP}\times\text{FN}\right)}{\left(\text{TP}+\text{FP}\right)\left(\text{FP}+\text{TN}\right)+\left(\text{TP}+\text{FN}\right)\left(\text{FN}+\text{TN}\right)},
\end{equation}
see, for example, \cite{chicco21}.
For the first method, we obtain 
$\kappa_{\text{local}} = 0.66$, 
while for the second $\kappa_{\text{global}}=0.62$. 
Both values put in evidence a good agreement and accuracy in predictions 
for both the methods.

Now, we aim to evaluate the performance of the combination of the two 
methods, with the purpose of improving predictions and, at the same time, 
reducing the variance in the predictions and, thus, enhancing their stability.
Table \ref{t:comp1} compares the reliability of the predictions of the two models, averaging over the outcomes of the 10--fold cross--validation 
performed for the optimal tuning parameters. 
It is important to remark that
when the outcomes for both models are either 
0 (TN or FN)
or 1 (TP or FP),  
the amount of correct prediction is
$92.2\%$ (negative predictive value) 
and $81\%$ (positive predictive value), respectively.
Furthermore the total proportion of wrong negative outcomes is $2.9\%$.

\begin{table}
	\begin{center}
		\begin{tabular}{ |c||c|c|c|c| }   	\hline
	Local (row) vs. Global (col.)& TP(\%) & FP(\%) & FN(\%) & TN(\%) \\
	\hline
	TP(\%) & 42.1 &  & 2.5 &  \\
	\hline
	FP(\%) &  & 9.7 &  & 1.9 \\
	\hline
	FN(\%) & 2.5 &  & 2.9   &  \\
	\hline
	TN(\%) &  & 4.1 &  & 34.3\\
	\hline	
		\end{tabular}
		\caption{Performance of the outcomes for the joint methods.
Empty cells correspond to impossible combinations of outcomes.
\label{t:comp1}}
	\end{center}
\end{table}

Table \ref{t:comp2} investigates on the accuracy of the predictions. 
Even if $11\%$ of the predictions are at odds, and thus they 
should be directly marked as unreliable, 
$86\%$ (joint accuracy) predictions are valid when both agree. 

\begin{table}
	\begin{center}
		\begin{tabular}{ |c||c|c| }   	\hline
Local (row) vs. Global (col.)& correct predictions (\%) & 
                               wrong predictions (\%) \\
			\hline
			correct predictions (\%)& 76.4 &  6.6 \\
			\hline
			wrong predictions (\%)  &  4.4 &  12.6 \\
			\hline	
		\end{tabular}
		\caption{
                 Correct vs. wrong predictions (joint models).
\label{t:comp2}}
	\end{center}
\end{table}

\section{Conclusions}
\label{s:conclusioni} 
\par\noindent
Raman spectra of tumoral and healthy genomic DNA 
have been collected by analyzing aqueous DNA droplets 
deposited onto silver--coated silicon nanowires. 
We used, respectively, the human melanoma cell line SK--MEL--28 and 
the human immortalized keratinocyte HaCaT as tumoral and healthy 
samples. 

Pre--processed spectra were analyzed by means of two different 
techniques: a PCA based algorithm powered with linear regression 
and a purely geometric algorithm were 
devised to predict the tumoral or healthy origin of each spectra. 
Both algorithms achieve very high accuracy, close to $90$\%.
We also checked accuracy by means of the
Cohen's
$\kappa$ statistic and for both algorithms we found 
very good values of the Cohen's $\kappa$ index -- larger than $0.6$.

On one hand the PCA approach, rather standard in analyzing Raman 
measurements, aims at reducing the data set through the analysis 
of the covariance matrix. On the other hand, the geometric method 
that we implemented in this paper looks at the whole set of data 
and is based on the simple computation of the $\ell^2$ distance 
between spectra. The fact that both methods work nicely in 
discriminating healthy and tumoral molecules is a strong sign 
of the robustness of the indications provided by the Raman 
measurements. 

In conclusion, we developed a detailed statistical analysis which proves 
that SERS measurements
can be successfully used for efficient and fast cancer diagnostic 
applications. In particular, our innovative classification approach allows 
a rapid and unexpansive discrimination between healthy and tumoral 
genomic DNA, and represents a powerful alternative to the more complex 
and expensive DNA sequencing. Moreover, the proposed algorithms, 
implemented in commercial benchtop Raman spectrophotometers, could 
open the way to a completely automatic diagnostic analysis.

\begin{acknowledgments}
The research activity is funded by Regione Lazio within the project DIANA,
``DIAgnostic potential of disorder: development of an innovative 
NAnostructured platform for rapid, label--free and low--cost analysis 
of genomic DNA'', POR FESR Lazio 2014 -- 2020, 
Progetti Gruppi di ricerca call 2020, 
A0375-2020-36589, CUP B85F21001240002.

A.C. acknowledges the support of the Italian Minister of Foreign Affairs and
International Collaboration
(MAECI) under the Joint research project ``Scalable nano--plasmonic platform 
for differentiation
and drug response monitoring of organtropic metastatic cancer cells''
(US19GR07).
\end{acknowledgments}



\vfill\eject

\bibliographystyle{unsrt}
\bibliography{ccddllms-raman_tum_stat}

\begin{thebibliography}{10}

\bibitem{kksn2015}
K.~Kong, C.~Kendall, N.~Stone, and I.~Notingher.
\newblock Raman spectroscopy for medical diagnostics --- {F}rom in--vitro
  biofluid assays to in--vivo cancer detection.
\newblock {\em Advanced Drug Delivery Reviews}, 89:121--134, 2015.

\bibitem{lpppsb2021}
Z.~Liu, S.~Parida, R.~Prasad, R.~Pandeya, D.~Sharma, and I.~Barman.
\newblock Vibrational spectroscopy for decoding cancer microbiota interactions:
  Current evidence and future perspective.
\newblock {\em Seminars in Cancer Biology}, 2021.

\bibitem{mlpmpbllc2021}
V.~Mussi, M.~Ledda, D.~Polese, L.~Maiolo, D.~Paria, I.~Barman, M.G. Lolli,
  A.~Lisi, and A.~Convertino.
\newblock Silver--coated silicon nanowire platform discriminates genomic {DNA}
  from normal and malignant human epithelial cells using label--free raman
  spctroscopy.
\newblock {\em Material Science \& Engineering C}, 122:111951, 2021.

\bibitem{mrr2007}
Z.~Movasaghi, S.~Rehman, and I.U. Rehman.
\newblock {Raman Spectroscopy of Biological Tissues}.
\newblock {\em Applied Spectroscopy Reviews}, 42:493--541, 2007.

\bibitem{tmrr2015}
A.C.S. Talari, Z.~Movasaghi, S.~Rehman, and I.U. Rehman.
\newblock {Raman Spectroscopy of Biological Tissues}.
\newblock {\em Applied Spectroscopy Reviews}, 50:46--111, 2015.

\bibitem{psp2003}
R.~Petry, M.~Schmitt, and J.~Popp.
\newblock Raman spectroscopy -- a prospective tool in the life sciences.
\newblock {\em Chemphyschem: a European journal of chemical physics and
  physical chemistry}, 4:14--30, 2003.

\bibitem{fhm1974}
M.~Fleischmann, P.J. Hendra, and A.J. McQuillan.
\newblock Raman spectra of pyridine adsorbed at a silver electrode.
\newblock {\em Chemical Physics Letters}, 26:163--166, 1974.

\bibitem{hmd2005}
C.L. Haynes, A.D. McFarland, and R.P.~Van Duyne.
\newblock Surface--{E}nhanced {R}aman {S}pectroscopy.
\newblock {\em Analytical Chemistry}, 77:338A--346A, 2005.

\bibitem{sdsd2008}
P.L. Stiles, J.A. Dieringer, N.C. Shah, and R.P.~Van Duyne.
\newblock Surface--{E}nhanced {R}aman {S}pectroscopy.
\newblock {\em Annual Review of Analytical Chemistry}, 1:601--626, 2008.

\bibitem{cmm2016}
A.~Convertino, V.~Mussi, and L.~Maiolo.
\newblock Disordered array of {A}u covered {S}ilicon nanowires for {SERS}
  biosensing combined with electrochemical detection.
\newblock {\em Scientific Reports}, 6:25099, 2016.

\bibitem{cmmllbfrl2018}
A.~Convertino, V.~Mussi, L.~Maiolo, M.~Ledda, M.G. Lolli, F.A. Bovino,
  G.~Fortunato, M.~Rocchia, and A.~Lisi.
\newblock Array of disordered silicon nanowires coated by a gold film for
  combined {NIR} photothermal treatment of cancer cells and {R}aman monitoring
  of the process evolution.
\newblock {\em Nanotechnology}, 29:415102, 2018.

\bibitem{zwlazc2008}
B.~Zhang, H.~Wang, L.~Lu, K.~Ai, G.~Zhang, and X.~Cheng.
\newblock Large--{A}rea {S}ilver--{C}oated {S}ilicon {N}anowire {A}rrays for
  {M}olecular {S}ensing {U}sing {S}urface--{E}nhanced {R}aman {S}pectroscopy.
\newblock {\em Advanced functional materials}, 18:2348--2355, 2008.

\bibitem{gbcspb2009}
E.~Galopin, J.~Barbillat, Y.~Coffinier, S.~Szunerits, G.~Patriarche, and
  R.~Boukherroub.
\newblock Silicon {N}anowires coated with {S}ilver {N}anostructures as
  {U}ltrasensitive {I}nterfaces for {S}urface--{E}nhanced {R}aman
  {S}pectroscopy.
\newblock {\em ACS Appl.~Mater.~Interfaces}, 7:1396--1403, 2009.

\bibitem{zfzszwl2010}
M.-L. Zhang, X.~Fan, H.-W. Zhou, M.-W. Shao, J.A. Zapien, N.-B. Wong, and S.-T.
  Lee.
\newblock A {H}igh--{E}fficiency {S}urface--{E}nhanced {R}aman {S}cattering
  {S}ubstrate {B}ased on {S}ilicon {N}anowires {A}rray {D}ecorated with
  {S}ilver {N}anoparticles.
\newblock {\em J.~Phys.~Chem.~C}, 114:1969--1975, 2010.

\bibitem{pcmmb2021}
D.~Paria, A.~Convertino, V.~Mussi, L.~Maiolo, and I.~Barman.
\newblock Silver--{C}oated {D}isordered {S}ilicon {N}anowires {P}rovide
  {H}ighly {S}ensitive {L}abel--{F}ree {G}lycated {A}lbumin {D}etection through
  {M}olecular {T}rapping and {P}lasmonic {H}otspot {F}ormation.
\newblock {\em Advanced healthcare materials}, 10:2001110, 2021.

\bibitem{shb2012}
M.S. Schmidt, J.~H\"ubner, and A.~Boisen.
\newblock Large {A}rea {F}abrication of {L}eaning {S}ilicon {N}anopillars for
  {S}urface {E}nhanced {R}aman {S}pectroscopy.
\newblock {\em Advanced materials}, 24:OP11--OP18, 2012.

\bibitem{wlhwgkhoe2016}
C.E.M. Weber, C.~Luo, A.~Hotz-Wagenblatt, A.~Gardyan, T.~Korda{\ss},
  T.~Holland-Letz, Wi. Osen, and S.B. Eichm{\"u}ller.
\newblock {miR--339--3p {I}s a {T}umor {S}uppressor in {M}elanoma}.
\newblock {\em Cancer research}, 76:3562--3571, 2016.

\bibitem{scf2009}
M.R. Stratton, P.J. Campbell, and P.A. Futreal.
\newblock The cancer genome.
\newblock {\em Nature}, 458:719--724, 2009.

\bibitem{SG64}
A.~Savitzky and M.J.E. Golay.
\newblock {Smoothing and Differentiation of Data by Simplified LeastSquares
  Procedures}.
\newblock {\em Anal. Chem.}, 36:1627--1639, 1964.

\bibitem{ZK13}
B.~Zimmermann and A.~Kohler.
\newblock Optimizing {S}avitzky--{G}olay parameters for improving spectral
  resolution and quantification in infrared spectroscopy.
\newblock {\em Appl. Spectrosc.}, 67(8):892--902, 2013.

\bibitem{htw2015}
T.~Hastie, R.~Tibshirani, and M.~Wainwright.
\newblock {\em {Statistical Learning with Sparsity: The Lasso and
  Generalizations}}.
\newblock CRC Press, Routledge, 2015.

\bibitem{Jackson91}
J.E. Jackson.
\newblock {\em A User's Guide to Principal Components.}
\newblock Wiley, 1991.

\bibitem{HTF09}
T.~Hastie, R.~Tibshirani, and J.~Friedman.
\newblock {\em The elements of statistical learning}.
\newblock Springer, 2009.

\bibitem{Fawcett2006}
T.~Fawcett.
\newblock An introduction to {ROC} analysis.
\newblock {\em Pattern Recognition Letters}, 28:861--874, 2006.

\bibitem{chicco21}
D.~Chicco, M.J. Warrensand, and G.~Jurman.
\newblock The {M}atthews correlation coefficient ({MCC}) is more informative
  than {C}ohen's {K}appa and {B}rier score in binary classification assessment.
\newblock {\em Access}, 9:78368--78381, 2021.

\end{thebibliography}

\end{document}